\documentclass[prd,showpacs,preprintnumbers,superscriptaddress,floatfix,twocolumn,amsmath,amssymb]{revtex4}
\usepackage{graphicx}% Include figure files
\usepackage{dcolumn}% Align table columns on decimal point
\usepackage{bm}% bold math

\newcommand{\tires}{\affiliation{TIRES: Center of Innovative Technologies for Signal Detection
and Processing, University of Bari, Italy}}
\newcommand{\fisica}{\affiliation{Physics Department, University of Bari, Italy}}
\newcommand{\infn}{\affiliation{Istituto Nazionale di Fisica Nucleare, Sezione di Bari}}
\newcommand{\deto}{\affiliation{D.E.T.O., University of Bari, Italy}}
\newcommand{\neuro}{\affiliation{Department of Neurological and Psychiatric Sciences,University of Bari, Italy}}
\newcommand{\dfboston}{\affiliation{Center for Polymer Studies and Department of Physics, Boston University, Boston,
  Massachusetts}}
\newcommand{\hms}{\affiliation{Beth Israel Deaconess Medical Center, Harvard Medical School, Boston Massachusetts}}

\begin{document}
\preprint{BA-TH 470}
\title{Steady-state visual evoked potentials and phase synchronization in migraine}
\date{\today}
\author{L. Angelini} \tires \fisica \infn
\author{M. De Tommaso} \tires \neuro
\author{M. Guido} \tires
\author{K. Hu} \dfboston \hms
\author{P. Ch. Ivanov}\dfboston \hms
\author{D. Marinazzo} \tires
\author{G. Nardulli} \tires \fisica \infn
\author{L. Nitti} \tires \deto \infn
\author{M. Pellicoro} \tires \fisica \infn
\author{C. Pierro} \tires
\author{S. Stramaglia} \tires \fisica \infn
\begin{abstract}

We investigate phase synchronization in EEG recordings from migraine patients.
We use the analytic signal technique, based on the Hilbert transform, and find
that migraine brains are characterized by enhanced alpha band phase
synchronization in presence of visual stimuli. Our findings show that migraine
patients have an overactive regulatory mechanism that renders them more
sensitive to external stimuli.
 \end{abstract}
  \maketitle

 Phase synchronization was introduced for coupled chaotic systems by
Rosenblum et al. \cite{Rosenblum1} and has been confirmed experimentally \cite{parlitz}.
This concept, introduced in the field of nonlinear dynamics, provides a measure of
synchronization alternative to conventional linear approaches. It may be useful for
biological time series, in particular to the study of electroencephalographic (EEG)
signals, where synchronization phenomena are expected to play a major role for
establishing the communication between different regions of the brain \cite{gray}.
\begin{figure}[ht]
\begin{center}
\includegraphics[width=7.0cm]{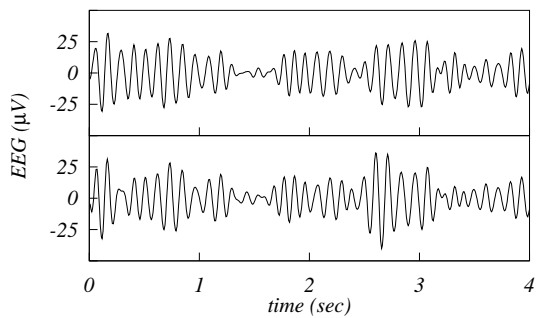}
\end{center}
\caption{\small Examples of EEG signals; data are taken from a migraine patient subject
to 9 Hz flash stimulation, and correspond to two frontal electrodes (F1 and F2). The full
record is 40 sec long, only a 4 sec segment is shown. The signals are filtered in the
alpha band.} \label{f1}
\end{figure}

\begin{figure}[ht]
\begin{center}
\includegraphics[width=7.0cm]{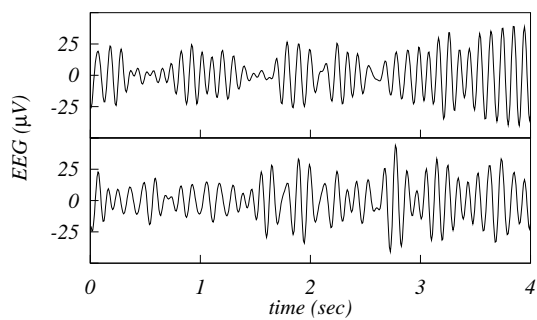}
\end{center}
\caption{\small Examples of spontaneous EEG; data, filtered in the alpha band, are taken
from the same migraine patient as in  fig. \ref{f1}, and correspond to F1 and F2
electrodes.} \label{f2}
\end{figure}
Migraine is an incapacitating disorder of neurovascular origin, which consists of attacks
of headache, accompanied by autonomic and possibly neurological symptoms. It is estimated
that in the USA, $5\%$ of the general population suffer at least 18 days of migraine a
year, and more than $1\%$ have at least one day of migraine a week \cite{Goadsby}. In
spite of a lot of research, there are still many unresolved issues in the pathophysiology
of migraine. There is a tendency to believe that migraine starts with an underlying
central nervous system disorder, which, when triggered by various stimuli, sets off a
chain of neurologic and biochemical events, some of which subsequently affect the brain's
vascular system. No experimental model fully explains the migraine process \cite{ihs}. A
wide range of events and conditions can alter conditions in the brain that bring on nerve
excitation and trigger migraines. They include emotional stress, intense physical
exertion, abrupt weather changes, flickering lights, and many others. The question we
address here is: how does the response of migraine patients to such events differs from
those of healthy persons? To address this problem, we investigate synchronization
phenomena in Electroencephalograms (EEGs) recorded from migraine patients in presence of
repetitive visual stimuli (steady-state visual evoked potentials, SVEPs \cite{silver}),
and study how synchronization between different brain regions varies in presence of
external stimuli (i.e., while brain is processing external information). We find that
migraine brains show increased alpha band phase synchronization, while healthy persons
show a decreased one. Our results suggest that migraine patients have an overactive
regulatory mechanism, prone to instability, which renders them more sensitive to
environmental factors.

\par Our data are as follows. EEG is recorded from fifteen
patients affected by  migraine without aura \cite{aura}, in presence of visual stimuli.
During the acquisition, flash stimuli are presented to the subjects repetitively at a
rate of 3-6-9-12-15-18-21-24-27 Hz. The mean age of patients is 38.7 years (range 24-48
years). Each frequency of stimulation is delivered by a flash settled at a luminance of
0.2 joules for at least 20 seconds; an interval of 20 seconds is interposed between the
different trains of stimulation. EEG data are recorded by 18 scalp electrodes, placed
according to the International 10-20 system, referred to CZ derivation. Impedance is
settled below 5 $K\Omega$, EEG is digitally filtered off line by means of a digital
filter with a band-pass of 0.3-30 Hz; the sampling rate is 128 Hz. Examples of EEG
signals are shown in fig. \ref{f1}; also spontaneous EEG (i.e. in the absence of stimuli)
is recorded for all patients, see fig. \ref{f2}. All patients are in the inter-ictal
state, the time from the end of the last attack being at least 72 hours. Moreover, EEG
data from fifteen healthy subjects (ages ranging from 22 to 45 years) are measured so as
to have a control group.

We recall how to detect $n:m$ phase synchronization in noisy scalar signals
\cite{tass98}. Based on general theorems on analytical functions the following relation
holds
\begin{equation}
{\rm Im}\, \zeta(t)=\frac 1 \pi
P.V.\int_{-\infty}^{+\infty}\frac{{\rm
Re}\,\zeta(\tau)}{t-\tau}\,d\tau\ .\label{ht}\end{equation}

This equation is known as the Hilbert transform and it is used to form, starting from a
signal $s(t)={\rm Re}\zeta(t)$, the analytic signal
 $\zeta(t)=A(t)e^{i\phi(t)}$,
 where $A(t)=\sqrt{s^2(t)+\tilde s^2(t)}$, and
 $\tilde s(t)={\rm Im} \zeta(t)$.
 To control the possible synchronization of two signals $s_1(t)$,
 $s_2(t)$ the following procedure is applied: the
instantaneous phases $\phi_1(t)$ and $\phi_2(t)$ are computed and the so called {\it
generalized phase differences}
\begin{equation} \Phi_{n,m}(t)=\left[m \phi_1(t)\,-\,n\phi_2(t)\right]_{mod
2\pi},\end{equation}  with $n,m$ integers, are evaluated \cite{wavelet}. Phase
synchronization is characterized by the appearance of peaks in the distribution of
$\Phi_{n,m}$ and quantified by comparing the actual distribution with a uniform one in
the following way. The $n:m$ synchronization index of $s_1$ and $s_2$ is defined as
$\rho_{n,m}=[S_{max}-S_{n,m}]/S_{max}$, where $S_{n,m}$ is the Shannon entropy of the
distribution of $\Phi_{n,m}$ and $S_{max}$ is the entropy of the uniform distribution; in
the case at hand only $1:1$ synchronization leads to interesting results.

\begin{figure}[ht]
\begin{center}
\includegraphics[width=8.5cm]{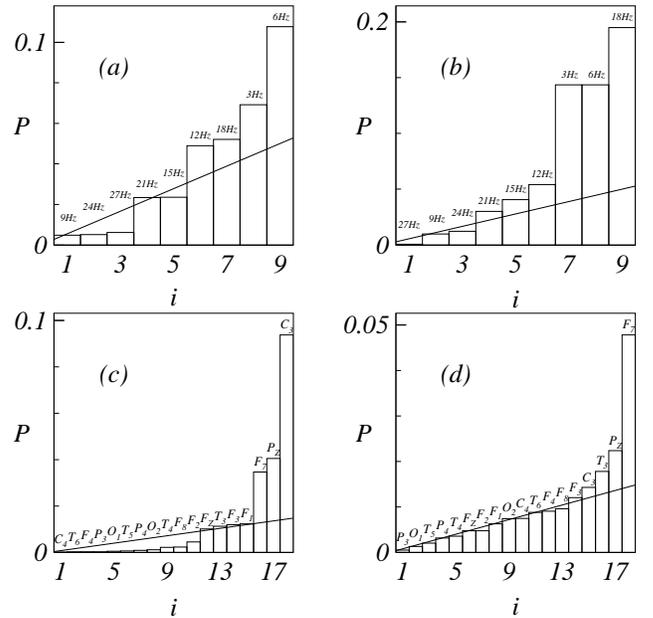}
\end{center}
\caption{\small (a) The FDR method (see the note \cite{fdr1}) is applied to select the
stimuli frequencies separating patients and controls according to $\Gamma$ values. The
vertical axis represent the probability that the thirty $\Gamma$ values of patients and
controls were drawn from the same distribution, according to the paired t-test.
Frequencies 9-24-27 Hz are selected with false positive rate $0.05$. (b) The probability
that the thirty $\Gamma$ values of patients and controls were drawn from the same
distribution is now evaluated according to the Wilcoxon rank sum test. FDR selects
frequencies 9-24-27 Hz with false positive rate $0.05$.(c) The FDR method is applied to
select separating electrodes, for 9 Hz flash stimuli. The vertical axis represent the
probability that the thirty $\Gamma_s$ values of patients and controls, for each sensor
$s$, were drawn from the same distribution, according to the paired t-test. Eleven
electrodes, out of eighteen, are selected. The labels for electrodes correspond to
International 10-20 system. (d) As in (c) for 24 Hz flash stimuli. Thirteen electrodes
are selected.  } \label{f3}
\end{figure}

\par Let us now turn to describe our findings. The EEG signals are filtered in
the alpha band (8-12.5 Hz) \cite{nota} and the synchronization index above described is
evaluated for all pairs of electrodes, for all thirty subjects and for all frequencies of
the flash stimuli. These indexes are subsequently averaged over all the possible pairs of
sensors, for each subject both in presence of stimuli and in spontaneous conditions.
These mean values do not separate patients from healthy subjects; what emerges as
correlated with the migraine pathology is the ratio $\gamma=\rho_{1,1}^f/
\rho_{1,1}^{sp}$, where $\rho_{1,1}^f$
 is the mean phase synchronization in presence of flash stimuli, whereas
$\rho_{1,1}^{sp}$ is the mean spontaneous phase synchronization. This ratio measures how
phase synchronization varies, in the presence of the stimuli, with respect to basal
conditions, i.e. the neat effect of the stimulus. Our supervised analysis (hypothesis
testing) shows that the index $\Gamma = \ln \left(\gamma\right)$ \cite{log}
 separates the class of patients and the class of controls for stimulus frequencies of
9,24,27 Hz. For each of the 9 flash stimuli frequencies $\omega$,  we apply the paired
t-test to evaluate the probability $P_\omega$ that indexes ${\Gamma}s$ were drawn from
the same distribution (the null hypothesis); in seven cases out of nine this probability
is less than 0.05, the standard value used in literature to reject the null hypothesis.
However, here we deal with multiple comparisons. To control the number of false
positives, we use the false discovery rate (FDR) method \cite{fdr}. This procedure
\cite{fdr1} selects the stimuli frequencies 9-24-27 Hz as separating patients from
controls (see fig.\ref{f3}a), with the expected fraction of false positive $0.05$. The
same frequencies (9-24-27 Hz) are selected by use of the standard Bonferroni correction
for multiple comparisons \cite{bonferroni} as well as by FDR if probabilities are
evaluated by the non-parametric Wilcoxon rank sum test (see fig.\ref{f3}b).
\begin{figure}[ht]
\begin{center}
\includegraphics[width=6.5cm]{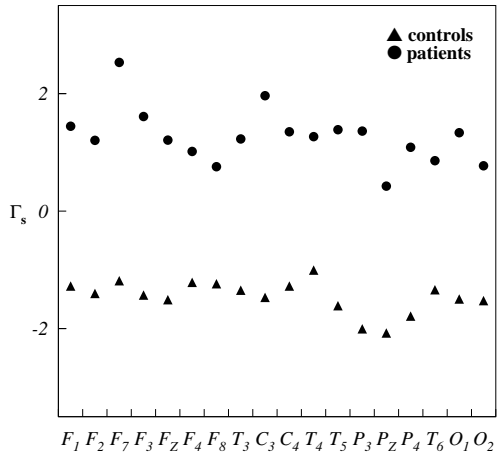}
\end{center}
\caption{\small In the case of 24 Hz stimuli, the mean of $\Gamma_s$ (over patients and
over controls) is represented for all the eighteen electrodes. On the average, phase
synchronization increases for patients and decreases for controls.}\label{f4}
\end{figure}

A topographic analysis is also performed, in order to check whether this phenomenon is
localized in some cortex region. We evaluate, for each sensor $s$, $\Gamma_s = \ln (
\langle \rho_{1,1}^f \rangle_s / \langle \rho_{1,1}^{sp}\rangle_s )$, where $\langle
\cdot \rangle_s$ means averaging only over the pairs where $s$ is one the two sensors.
For each frequency of stimuli, we apply FDR method to select, among the eighteen
electrodes, those separating patients from controls according to their $\Gamma_s$
\cite{fdr2}. The results are depicted in Figures \ref{f3}c (9 Hz case) and \ref{f3}d (24
Hz): eleven electrodes are recognized as separating in the case of 9 Hz stimuli and
thirteen in the case of 24 Hz; no electrode is found to be individually separating when
27 Hz stimuli are considered. Since separating electrodes from all the regions of the
cortex (frontal, parietal, central, temporal and occipital) are found, it follows that
the phenomenon here described is extended over all the cortex, not being localized in a
limited region. Its diffuse nature suggests that genuine spatial synchronization
\cite{srinivasan} is here involved; indeed, volume conduction effects \cite{menon} would
induce  spatially more localized change.

\begin{figure}[ht]
\begin{center}
\includegraphics[width=7.0cm]{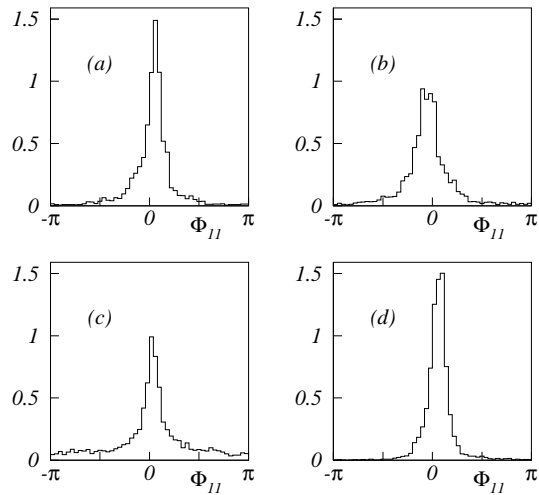}
\end{center}
\caption{\small Histogram of $\Phi_{1,1}$ for the pair T3-T5. (a) A healthy person
without stimuli. (b) The same healthy person in presence of 9 Hz stimuli. (c) A patient
without stimuli. (d) The same patient in presence of 9 Hz stimuli. } \label{f5}
\end{figure}
\par
\begin{figure}[ht]
\begin{center}
\includegraphics[width=6.3cm]{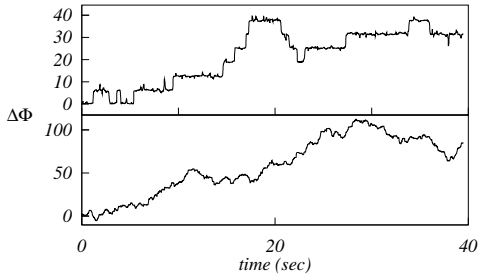}
\end{center}
\caption{\small Time evolution of $\Delta \Phi$ (i.e., the phase difference $\phi_1 -
\phi_2$, without wrapping in the interval $[-\pi,\pi]$) for the pair F1-F2 in a migraine
patient (the same patient as in fig. \ref{f2}). Top: in presence of 9 Hz stimuli. Bottom:
without stimulation. } \label{f6}
\end{figure}

\begin{figure}[ht]
\begin{center}
\includegraphics[width=6.3cm]{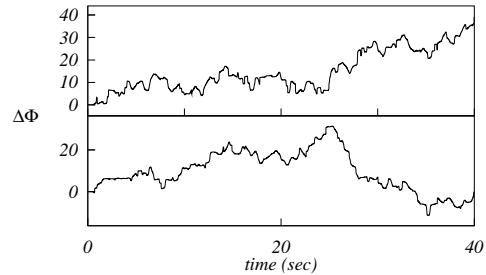}
\end{center}
\caption{\small Time evolution of $\Delta \Phi$ for pair F1-F2 for a control. Top: in
presence of 9 Hz stimuli. Bottom: without stimulation.} \label{f7}
\end{figure}

 Our data show that, for patients, the mean phase synchronization increases
in presence of visual stimuli, whereas it decreases in controls. For example, in the case
of 24 Hz stimuli, and for all the sensors, the mean value (over subjects) of $\Gamma_s$
is shown in fig. 4: hyper phase synchronization is observed in patients, whereas healthy
subjects show a reduced phase synchronization. Similar patterns occur for 9 and 27 Hz
stimuli. In fig. 5 the histograms of $\Phi_{1,1}$, corresponding to electrodes T3 and T5,
are shown for a migraine patient and for a control, both under stimulation and
spontaneously. The distribution, when stimuli are delivered, broadens for the healthy
person while becoming more peaked for the patient. This behavior is further illustrated
in fig.\ref{f6}, where the time evolution of the phase difference between two sensors is
depicted for a migraine patient, both subject to stimuli and in spontaneous conditions.
In presence of flash phase locking, in the two signals, is observed for time segments
several seconds long; no such locking is observed in the spontaneous case. Phase
difference curves, for a control, are drawn in fig.\ref{f7}. It is worth stressing that
this phenomenon is not mined if coherence is used to measure synchronization: considering
the linear index obtained by integration of the coherence function (normalized amplitude
of the cross spectrum of the two time series \cite{quiroga}) in the alpha band, the
corresponding $\Gamma$ and $\Gamma_s$ quantities do not lead to separation between
patients and controls for any frequency of stimulation.
We show that migraineurs are characterized by alpha band hyper-synchronization in
presence of visual stimuli. We also show how this varies with the frequency of the flash,
and present a topographic analysis where separating electrodes are recognized. Whilst it
is comprehensible that 9 Hz stimuli might cause hyper-synchronization in the alpha band
(8-12.5 Hz), in order to figure how 24-27 Hz stimuli may act on alpha oscillations we
observe that brain is a nonlinear system, and sub-harmonics of 24-27 Hz fall in the alpha
band: stimulation in the 24-27 band may cause hyper-synchronization through their
sub-harmonics. However a similar behavior is not observed for other frequencies with
sub-harmonics in the alpha band, like $18$ Hz: further investigation is needed to clarify
this aspect of the phenomenon. It will be also interesting to investigate the response of
migraine patients with aura. Our results are consistent with current theories about the
role of subcortical structures in migraine. Since brainstem is active in migraine
\cite{bahara}, it has been proposed, as a unifying concept of migraine, that brainstem
regions concerned with neural mechanism of synchrony are dysfunctional \cite{go}. Cortex,
in migraine brains, is thus misled by a dysfunctional gating system; normal light is
unpleasant, normal sound uncomfortable and, probably, normal pulsing of vessels felt as
pain. On the mathematical side, our results confirm the usefulness of the analytic signal
technique to study physiological time series.


\begin{thebibliography}{99}
\bibitem{nota} Alpha rhythms are associated with quiet wakefulness with eyes closed.
Both filtering the signals in the other
physiological bands (beta, theta, etc.) and filtering around the fundamental frequency of
the luminous stimuli, do not lead to significative differences in the measures of
synchronization of patients and healthy subjects.
\bibitem{log} We consider $\ln \left(\gamma\right)$ instead of $\gamma$ because the
distribution of $\ln \left(\gamma\right)$ values is found to be Gaussian, thus justifying
use of t-test.
\bibitem{fdr1} FDR works in the
following way: the $N=9$ frequencies are ordered according to
their increasing $P_\omega$ values, and a parameter $q$, which
controls the fraction of false positive, is set to $0.05$. The
index $j$ is identified as the largest such that for all $i\le j$
we have $P_{\omega_i}\le \frac{i}{N}\,\frac{q}{c(N)}$, $c(N)$
being a constant described below. The null hypothesis is rejected
for those frequencies whose index $i$ is such that $i\le j$. The
choice of the constant $c(N)$ depends on the assumptions about the
joint distribution of the P-values corresponding to different
frequencies: $c(N)$ is set to $1$ under the assumptions of
independence of the measures at different frequencies, and this
choice we make in the present case.
\bibitem{fdr2} In the case of correlated P-values the value
$c(N)=\sum_{i=1}^N{1/ i}$ is to be used in FDR. Since P-values from electrodes are not
strictly independent, we adopt, in this case, this choice with $N=18$.
\end{thebibliography}
\end{document}